\documentclass[prd,superscriptaddress,amsfonts,amssymb,amsmath,showpacs,twocolumn,10 pt,superscriptaddress,nofootinbib]{revtex4-2}
\usepackage{bm}
\usepackage{amsfonts}
\usepackage{latexsym}
\usepackage[latin1]{inputenc}
\usepackage{graphicx}
\usepackage{amsmath,eqnarray}
\usepackage{palatino}
\usepackage{mathpazo}
\usepackage{textcomp}
\usepackage{float}
\usepackage{subcaption}
\usepackage{booktabs}
\usepackage{dcolumn}
\usepackage{hyperref}
\usepackage{mathtools}
\hypersetup{colorlinks,citecolor=red}
\usepackage{footnote}
\usepackage{amsmath}
\usepackage{xcolor}
\usepackage{orcidlink}
\usepackage{ragged2e}
\usepackage{amsmath}
\usepackage{listings}
\usepackage{graphicx}
\def\jnl@style{\it}
\def\aaref@jnl#1{{\jnl@style#1}}

\def\aaref@jnl#1{{\jnl@style#1}}

\def\aj{\aaref@jnl{AJ}}                   
\def\apj{\aaref@jnl{ApJ}}                 
\def\apjl{\aaref@jnl{ApJ}}                
\def\apjs{\aaref@jnl{ApJS}}               
\def\apss{\aaref@jnl{Ap\&SS}}             
\def\aap{\aaref@jnl{A\&A}}                
\def\aapr{\aaref@jnl{A\&A~Rev.}}          
\def\aaps{\aaref@jnl{A\&AS}}              
\def\mnras{\aaref@jnl{Mon.~Not.~Roy.~Astron.~Soc.}}             
\def\prd{\aaref@jnl{Phys.~Rev.~D}}        
\def\prc{\aaref@jnl{Phys.~Rev.~C}}  
\def\prl{\aaref@jnl{Phys.~Rev.~Lett.}}    
\def\qjras{\aaref@jnl{QJRAS}}             
\def\skytel{\aaref@jnl{S\&T}}             
\def\ssr{\aaref@jnl{Space~Sci.~Rev.}}     
\def\zap{\aaref@jnl{ZAp}}                 
\def\nat{\aaref@jnl{Nature}}              
\def\aplett{\aaref@jnl{Astrophys.~Lett.}} 
\def\apspr{\aaref@jnl{Astrophys.~Space~Phys.~Res.}} 
\def\physrep{\aaref@jnl{Phys.~Rep.}}      
\def\physscr{\aaref@jnl{Phys.~Scr}}       
\def\commat{\aaref@jnl{Comm.~Math.~Phys.}}              
\def\science{\aaref@jnl{Science}}               
\def\cqg{\aaref@jnl{Classical Quant.~Grav.}}            
\def\jpcs{\aaref@jnl{JPCS}}                                     
\def\ijmpd{\aaref@jnl{Int.~J.~Mod.~Phys.~D}}                    
\def\grg{\aaref@jnl{Gen.~Relat.~Gravit.}}               
\def\rpp{\aaref@jnl{Rep.~Prog.~Phys.}}          
\def\npa{\aaref@jnl{Nucl.~Phys.~A}}        
\def\lrr{\aaref@jnl{Living Rev.~Rel.}}                   
\def\jcap{\aaref@jnl{J.~Cosmology Astropart.~Phys.}}    
\def\rmp{\aaref@jnl{Rev.~Mod.~Phys.}}   
\def\epjc{\aaref@jnl{Eur.~Phys.~J.~C}}

\allowdisplaybreaks[1]

\begin{document}

\color{black}

\author{Gaurav N. Gadbail\orcidlink{0000-0003-0684-9702}}
\email{gauravgadbail6@gmail.com}
\affiliation{Faculty of Symbiotic Systems Science, Fukushima University, Fukushima 960-1296, Japan.}

\author{Kazuharu Bamba \orcidlink{0000-0001-9720-8817}}
\email{bamba@sss.fukushima-u.ac.jp}
\affiliation{Faculty of Symbiotic Systems Science, Fukushima University, Fukushima 960-1296, Japan.}

\title{A sound-horizon-free measurement of the Hubble constant from DESI DR2 baryon acoustic oscillations using artificial neural networks}

\begin{abstract}
We present a model-independent, sound-horizon-free measurement of the Hubble constant $H_0$ using baryon acoustic oscillation tracers from the Dark Energy Spectroscopic Instrument Data Release 2. The function reconstructions are performed using the artificial neural network method, which is a completely data-driven approach that avoids the mild $\Lambda$CDM prior dependence. Our approach is based on the distance duality relation and combines three complementary observational probes, such as Type\,Ia supernovae, cosmic chronometer, and DESI DR2 BAO -- without requiring any knowledge of the sound horizon scale $r_d$ or any assumption about the absolute luminosity of SNe\,Ia. We obtain a joint constraint of  $H_0 = 71.5\pm2.2$\,km\,s$^{-1}$\,Mpc$^{-1}$ at 68\% confidence for 1000 bootstrap realisations and 4096 neurons, which is consistent with the TRGB result and the SH0ES measurement within $0.6\sigma$, consistent with the Planck 2020 result within $2\sigma$.  Our results favor a higher value of $H_0$ compared to the Planck CMB inference, adding independent support for the reality of the Hubble tension.


\end{abstract}

\maketitle


\section{Introduction}
 The Hubble ($H_0$) tension is the most fundamental problem in modern cosmology, which describes the present expansion rate of the universe. Despite decades of observational effort, the value of $H_0$ is still under intense debate, with measurements from different probes yielding results that are statistically inconsistent with each other. On one side of the tension, measurements based on the early universe consistently favour a lower value of $H_0$. The most precise of these comes from the Planck collaboration, whose analysis of the cosmic microwave background (CMB) anisotropies within the standard $\Lambda$CDM model yields $H_0 = 67.4 \pm 0.5$\,km\,s$^{-1}$\,Mpc$^{-1}$ \cite{Aghanim/2020}.  On the other side, measurements anchored in the local universe consistently prefer a higher value. The most prominent of these is the SH0ES result, which uses Type\,Ia supernovae (SNe\,Ia) calibrated via Cepheid variable stars to obtain $H_0 = 73.04 \pm 1.04$\,km\,s$^{-1}$\,Mpc$^{-1}$ \cite{Riess/2022}, in tension with Planck at the $5\sigma$--$6\sigma$ level. A third independent measurement using the Tip of the Red Giant Branch (TRGB) as a distance calibrator finds an intermediate value of $H_0 = 69.8 \pm 1.9$\,km\,s$^{-1}$\,Mpc$^{-1}$ \citep{Freedman2020}, consistent with both Planck and SH0ES within $2\sigma$. A key limitation of both the CMB and the local distance ladder approaches is that they rely on calibrations that introduce model dependence. The CMB measurement requires assuming $\Lambda$CDM to extrapolate from the acoustic scale at recombination to the present-day expansion rate, making the inferred $H_0$ sensitive to the assumed cosmological model. The local distance ladder requires anchoring the absolute luminosity of SNe\,Ia through Cepheid or TRGB calibrations, which carry their own systematic uncertainties, and have been the subject of ongoing debate \cite{Riess/2022,Freedman2021}. These dependencies motivate the development of completely independent, model free methods for measuring $H_0$ that bypass both the early-universe physics and the local calibration problem simultaneously.

Baryon acoustic oscillations (BAO) offer a promising path in this direction. BAO are the remnant imprint of sound waves that propagated through the primordial plasma before recombination, and they manifest as a preferred scale in the distribution of galaxies at $\sim 150$\,Mpc. As a standard ruler, BAO measurements provide precise constraints on cosmological distances across a wide range of redshifts. 
The Dark Energy Spectroscopic Instrument \cite{Levi2013} represents the state of the art in BAO surveys, covering over 14,200 square degrees of sky and targeting multiple galaxy tracers across the redshift range $0.1 < z < 4.2$. The recently released DESI Data Release 2 (DR2) \cite{Karim/2025a,Karim/2025b} provides BAO measurements of unprecedented precision from nine tracers, including one bright galaxy sample (BGS), three luminous red galaxy bins (LRG1, LRG2, LRG3), one combined LRG3$+$ELG1 bin, two emission-line galaxy bins (ELG1, ELG2), one quasar bin (QSO), and the Lyman-$\alpha$ forest (Ly$\alpha$), spanning the redshift range $0.295 \leq z \leq 2.330$.

An elegant method for extracting $H_0$ from BAO data along with SNe Ia and CC data without any sound horizon calibration was proposed in~\cite{Renzi2023} and subsequently applied to DESI DR1 BAO, SNe Ia, and CC data in~\cite{Guo2025}. The method combines three independent observational probes -- SNe\,Ia, cosmic chronometers (CC), and BAO -- through the distance duality relation, which holds in any metric theory of gravity. By forming a specific ratio of the three measured quantities, the sound horizon 
$r_d$ cancels exactly, the absolute magnitude of SNe\,Ia never appears, and the result depends only on directly observable distances and expansion rates. Using DESI DR1 BAO data and Gaussian Process (GP) regression for function reconstruction \cite{Guo2025}, obtained $H_0 = 68.4^{+1.0}_{-0.8}$\,km\,s$^{-1}$\,Mpc$^{-1}$, consistent with Planck within $1\sigma$ and in $4.3\sigma$ tension with SH0ES. Recent nonparametric studies of the expansion history using DESI BAO and SN\,Ia data have provided new insights into the Hubble tension \citep{Jiang2024}, motivating the model-independent analysis. Furthermore, CMB and Pantheon+SH0ES data have been used to investigate the ability of the $\tilde{\Lambda}$CDM model and its two-parameter extension, the $e\tilde{\Lambda}$CDM model (in which the critical indices $\delta_G$ and $\delta_\Lambda$ governing the redshift running of $G$ and $\Lambda$ are treated as independent parameters), to alleviate the $H_0$ tension in light of the recent DESI DR2 results~\cite{Zhang}.

In this work, we present a model-independent measurement of the Hubble constant $H_0$ using the latest BAO measurements from DESI DR2 \cite{Karim/2025a,Karim/2025b} via the artificial neural network (ANN) reconstruction method. In recent 
years, ANN-based methods have shown outstanding performance in solving cosmological problems in both accuracy and efficiency \cite{Wang2020, Fluri2019, George2018, Escamilla2020}. Our approach is based on the distance duality relation (DDR) and combines three complementary observational probes such as SNe\,Ia, CC, and BAO -- without requiring any knowledge of the sound horizon scale $r_d$ or any assumption about the absolute luminosity of SNe\,Ia. Basically, we extend the analysis in~\cite{Guo2025} in two important ways. First, we replace the GP reconstruction with the ANN reconstruction bootstrap approach in~\cite{Wang2020}, which is a completely data-driven approach that makes no assumption about the functional form of the reconstructed quantities and does not require a kernel assumption to perform its regression analysis, unlike the GP case\footnote{Both GP and ANN are powerful approaches for reconstructing smooth functions from sparse data. GP provides a natural probabilistic uncertainty estimate through its covariance structure, but requires a prior choice of kernel function that encodes assumptions about the smoothness and correlation length of the reconstructed quantity. In contrast, ANN requires no such kernel assumption and learns the functional form directly from the data, making it a more assumption-free reconstruction approach, and comparing their results offers a valuable consistency check on the reconstruction procedure.} \cite{Eoin}. Second, we apply the method to the newly released DESI DR2 BAO data, which provides seven independent anisotropic tracers compared to the five available in DR1, adding ELG1 at $z = 0.955$ and QSO at $z = 1.484$ as new independent redshift bins and significantly improving the precision of the existing measurements. We perform the analysis with two bootstrap configurations of $(N_{\rm boot},\, 
N_{\rm iter},\,N_{\rm neurons}) = (500,\,30000,\,4096)$ and 
$(1000,\,30000,\,4096)$ to verify the stability of our results, and we propagate all sources of uncertainty, including the SNe\,Ia calibration constant $\sigma_{a_B}$, ensuring a robust and complete 
propagation of all sources of uncertainty through the ANN reconstruction.

This paper is organised as follows. In Sec.~\ref{section 2}, we describe the theoretical framework method, the three observational datasets, and  the ANN reconstruction method. In Sec.~\ref{sec:H0}, we describe the $H_0$ posterior construction for each BAO redshift and joint constraint. In Sec.~\ref{section 4} we present our results at each BAO redshift and the joint constraint, in comparison with the literature, and discuss the implications for the Hubble tension. Sec.~\ref {section 5}, our findings are summarised.

\section{Methodology}\label{section 2}
\subsection{Method}
The Hubble constant $H_0$ describes the present expansion rate of the universe and appears in all cosmological distance measures. In this work, we determine $H_0$ in a fully model-independent manner by combining three complementary observational probes through the distance duality relation 
(DDR). The DDR states that in any metric theory of gravity, provided photons travel along null geodesics and their number is conserved. In the framework of DDR, two cosmological distance i.e., the luminosity distance $d_L(z)$ and the angular diameter distance $d_A(z)$ at the same redshift $z$ are connected as
\begin{equation}
    d_L(z) = (1+z)^2\,d_A(z).
    \label{eq:ddr}
\end{equation}
This relation has been extensively tested and confirmed 
observationally \citep{Cao2011, Cao2016, Qi2019, Liu2023, Zheng2020}. 
Combining Eq.~(\ref{eq:ddr}) with the definitions of the three observable quantities: $\Xi_{\rm SNIa}(z) = H_0\,d_L(z)$ is the unanchored luminosity distance from Type Ia supernovae, $H_{\rm CC}(z)$ is the Hubble parameter from cosmic chronometers, and $\Xi_{\rm BAO}(z) = H(z)\,d_A(z)$ is a sound-horizon-free combination from baryon acoustic oscillation measurements, the Hubble constant can be written as \citep{Guo2025}
\begin{equation}
    H_0 = \frac{\Xi_{\rm SNIa}(z)\, H_{\rm CC}(z)}
               {(1+z)^2\,\Xi_{\rm BAO}(z)},
    \label{eq:h0}
\end{equation}
Since $H_0$ appears in both $\Xi_{\rm SNIa}$ and $H_{\rm CC}$, the right-hand side of Eq.~(\ref{eq:h0}) is independent of $H_0$, making the approach genuinely self-consistent. Crucially, no cosmological model, no sound horizon calibration, and no assumption about the absolute magnitude of SNe\,Ia are required.

\subsection{Observational Datasets}
\label{sec:data}
\paragraph{Type Ia Supernovae (SNe Ia) - Pantheon$+$:} 
The SNe Ia data are taken from the Pantheon$+$SH0ES compilation \citep{Brout2022}, which includes 1701 light-curve measurements from 1550 distinct supernovae spanning the redshift range $z \in [0.001,\,2.26]$. We exclude supernovae at $z < 0.01$ to avoid peculiar-velocity contamination, and we exclude the 77 Cepheid calibrators to prevent any implicit $H_0$ prior from entering our analysis. After these cuts our working sample contains $\sim$1590 SNe Ia.

The observed B-band magnitude $m_B$ of each supernova is related to the unanchored luminosity distance through 
\begin{equation}
    \Xi_{\rm SNIa}(z) \equiv H_0\,d_L(z) 
    = 10^{\,0.2\,m_B + a_B},
    \label{eq:xi_snia}
\end{equation}
where the calibration constant $a_B = 0.71273 \pm 0.00176$ 
absorbs the degeneracy between $H_0$ and the absolute magnitude $M_B$ \citep{Riess2016}. The total uncertainty on $\Xi_{\rm SNIa}$ arise from both the photometric uncertainty $\sigma_{m_B}$ and the calibration uncertainty $\sigma_{a_B}$ as independent contributions:
\begin{equation}
    \sigma_{\Xi_{\rm SNIa}} = \Xi_{\rm SNIa}\cdot\ln(10)
    \cdot\sqrt{\bigl(0.2\,\sigma_{m_B}\bigr)^2 + \sigma_{a_B}^2}.
    \label{eq:sigma_xi_snia}
\end{equation}
 The inclusion of $\sigma_{a_B}$ in the above equation ensures that the calibration uncertainty is fully propagated into the final $H_0$ posterior, which is essential for a robust error budget.
 
\paragraph{Cosmic Chronometer (CC):} The Hubble parameter $H(z)$ is measured directly from the differential ages of passively evolving massive galaxies via the CC method \citep{Jimenez2002,CC1,CC2,CC3,CC4,CC5,CC6}:
\begin{equation}
    H(z) = -\frac{1}{1+z}\frac{\mathrm{d}z}{\mathrm{d}t}.
    \label{eq:cc}
\end{equation}
Since the redshift $z$ is measured spectroscopically with high 
accuracy, and the differential age evolution $\mathrm{d}t$ is derived from the spectral evolution of massive early-type galaxies that formed the bulk of their stellar mass at $z > 2$--$3$ without major subsequent star-formation episodes, this method is model-independent. We use 32 CC measurements that span $0.07 \leq z \leq 1.965$ \citep{Qi2023}.

\paragraph{DESI BAO DR2:}  We use BAO measurements from the second data release of DESI, which includes observations of galaxies and quasars \cite{Karim/2025b}, as well as Lyman-$\alpha$ tracers \cite{Karim/2025a}. These measurements cover both isotropic and anisotropic BAO constraints over $0.295 \leq z \leq 2.330$, divided into nine redshift bins \cite{Karim/2025a}. The DESI BAO DR2 measurements provide constraints that are expressed in terms of the transverse comoving distance $D_M/r_d$, the Hubble horizon $D_H/r_d$, and the angle-averaged distance $D_V(z)/r_d$, all normalized to the comoving sound horizon at the drag epoch, $r_d$. By forming the ratio of the transverse comoving distance and the Hubble horizon, the sound horizon cancels exactly and we get,
\begin{equation}
    \Xi_{\rm BAO}(z) \equiv H(z)\,d_A(z) 
    = \frac{c\,(D_M/r_d)}{(D_H/r_d)\,(1+z)}= \frac{c\,D_M}{D_H\,(1+z)},
    \label{eq:xi_bao}
\end{equation}
so that no CMB-based prior on $r_d$ is needed. The uncertainty 
on $\Xi_{\rm BAO}$ is propagated using the full covariance between $D_M/r_d$ and $D_H/r_d$, characterised by their correlation coefficient $\rho$:
\begin{equation}
    \rho = \frac{\mathrm{Cov}(D_M/r_d,\;D_H/r_d)}
                {\sigma_{D_M/r_d}\;\sigma_{D_H/r_d}},
    \label{rho_def}
\end{equation}
giving:
\begin{multline}
    \sigma^2_{\Xi_{\rm BAO}} = 
    \left(\frac{\partial\Xi}{\partial (D_M/r_d)}\sigma_{D_M/r_d}\right)^2
    + \left(\frac{\partial\Xi}{\partial (D_H/r_d)}\sigma_{D_H/r_d}\right)^2\\
    + 2\rho\,
      \frac{\partial\Xi}{\partial (D_M/r_d)}
      \frac{\partial\Xi}{\partial (D_H/r_d)}
      \sigma_{D_M/r_d}\sigma_{D_H/r_d}.
    \label{eq:sigma_xi_bao}
\end{multline}

In this work, we use seven anisotropic DESI DR2 tracers: three luminous red galaxy bins (LRG1 at $z=0.510$, LRG2 at $z=0.706$, LRG3 at $z=0.922$), two emission-line galaxy bins (ELG1 at $z=0.955$, ELG2 at $z=1.321$), one quasar bin (QSO at $z=1.484$), and the Lyman-$\alpha$ forest (Ly$\alpha$ at $z=2.330$). The ELG1 and QSO bins are new in DR2 and 
were not available in the DR1. From the full DESI DR2 BAO catalogue we exclude two data points. The first is the Bright Galaxy Sample (BGS) at $z=0.295$, which provides only an isotropic BAO measurement and does not report the individual $D_M/r_d$ and $D_H/r_d$ values required by Eq.~(\ref{eq:xi_bao}); without both transverse and line-of-sight distance ratios the sound-horizon-free combination $\Xi_{\rm BAO}$ cannot be constructed. The second excluded point is the combined LRG3$+$ELG1 bin at $z=0.934$, which overlaps in redshift with the individual LRG3 ($z=0.922$) and ELG1 ($z=0.955$) bins that we retain separately; including it alongside the individual bins would introduce a double-counting of the same galaxies and produce correlated data points in our joint posterior. The remaining seven tracers span the redshift range $0.510 \leq z \leq 2.330$ and constitute a set of independent BAO measurements suitable for our $H_0$ analysis.
\subsection{ANN Function Reconstruction}
\label{sec:ann}
The BAO measurements exist only at 7 discrete redshifts. To evaluate $\Xi_{\rm SNIa}(z)$ and $H_{\rm CC}(z)$ at exactly these redshifts, we reconstruct smooth continuous functions from the data using the ANN reconstruction method \cite{Wang2020}. This approach is completely data-driven and makes no assumption about the functional form of the reconstructed quantity or about the Gaussian nature of the observational uncertainties.
\subsubsection{Network Architecture and Training Procedure}
The network consists of:
\begin{itemize}
    \item An input layer accepting the scalar\footnote{A scalar is a single numerical value.} redshift $z$.
    \item One hidden layer with 4096 neurons and an Exponential Linear Unit (ELU) activation function \citep{Clevert2015}:
          \begin{equation}
              f(x) = \begin{cases} x, &  \mathrm{for}\,\,\, x > 0 \\ 
              \alpha(\mathrm{e}^x - 1), &  \mathrm{for}\,\,\,x \leq 0 \end{cases}
          \end{equation}
          with $\alpha = 1$.
    \item An output layer with 2 neurons returning the reconstructed function value and its uncertainty simultaneously.
    \item Batch normalisation\footnote{Batch normalisation standardises the distribution of layer activations within each training batch, stabilising the optimisation landscape and reducing sensitivity to the random initialisation of network weights. For the SNe\,Ia reconstruction, batch normalisation is deliberately disabled following \cite{Wang2020}, who found that it does not improve performance and can slightly degrade the smoothness of the reconstructed curve at high redshift where the data become sparse.} \cite{Ioffe2015} applied before the hidden layer for the CC reconstruction; disabled for the SNe\,Ia reconstruction \cite{Wang2020}. 
\end{itemize}

The choice of one hidden layer and 4096 neurons follows directly from the risk-minimisation procedure of \cite{Wang2020}. In their study, they evaluated 32 network architectures by systematically varying the number of hidden layers from one to four and the number of neurons per layer across eight values spanning the range $[128,\,16384]$. They found that as the number of neurons increases, the risk decreases first and then increases, reaching its minimum when the number of neurons is 4096. Therefore, we adopt a network containing 4096 neurons in the hidden layer as the optimal choice and apply it to the reconstruction.

The network is trained by minimising the mean absolute deviation loss function:
\begin{equation}
    \mathcal{L} = \frac{1}{mp}\sum_{i=1}^{m}\sum_{j=1}^{p}
    \bigl|\hat{Y}_{ij} - Y_{ij}\bigr|,
    \label{eq:loss}
\end{equation}
where $\hat{Y}$ is the network prediction, $Y$ contains the 
observed values and their uncertainties, $m$ is the batch size, and $p = 2$ is the number of output neurons. The Adam optimiser \citep{Kingma2014} is used with an initial learning rate of $0.01$, reduced to $10^{-5}$ via a cosine annealing schedule over 30\,000 training iterations. The batch size is set to half the number of data points in each dataset. Furthermore, we set the number of iterations to be 30\,000, which is large enough to ensure the loss function no longer decreases. We additionally tested $N_{\rm iter} = 20000$ and $N_{\rm iter} = 40000$ and found that all configurations yield fully consistent $H_0$ constraints within their uncertainties, confirming that the network has fully converged at our fiducial choice of $N_{\rm iter} = 30000$.
\subsubsection{Log-Space Reconstruction for SN Ia}
Since $\Xi_{\rm SNIa}$ spans several orders of magnitude across the redshift range $z \in [0.01,\,2.26]$, we perform the ANN reconstruction in $f(z)=\log_{10}(\Xi_{\rm SNIa})$ space for numerical stability. The log-space uncertainty is:
\begin{equation}
    \sigma_{f} = \sqrt{(0.2\,\sigma_{m_B})^2 + \sigma_{a_B}^2},
    \label{eq:sigma_logxi}
\end{equation}
 which correctly includes the $a_B$ calibration uncertainty. 
After reconstruction the results are converted back via 
$\Xi_{\rm SNIa} = 10^{\bar{f}}$ and $\sigma_{\Xi_{\rm SNIa}} = \Xi_{\rm SNIa}\cdot\ln(10)\cdot\sigma_{\bar{f}}$, where $\bar{f}$ and $\sigma_{\bar{f}}$ are the reconstructed log-space mean and standard deviation, respectively.

\subsection{Bootstrap Uncertainty Quantification}
\label{sec:bootstrap}
A single ANN training produces one reconstruction curve. To quantify the full propagation of observational uncertainties through the reconstruction, we follow the bootstrap procedure in~\cite{Wang2020} with $N_{\rm boot} = 500 $ and $1000$ realisations. The main idea is that we repeat the entire training process $N_{\rm boot}$ times, each time on a slightly different version of the data obtained by randomly perturbing each observation within its measurement uncertainty. This produces $N_{\rm boot}$ independent reconstruction curves, whose mean and spread at each redshift directly reflect the best estimate and the propagated uncertainty of the reconstructed function, respectively.

\begin{figure*}
    \centering
    \includegraphics[width=0.46\linewidth]{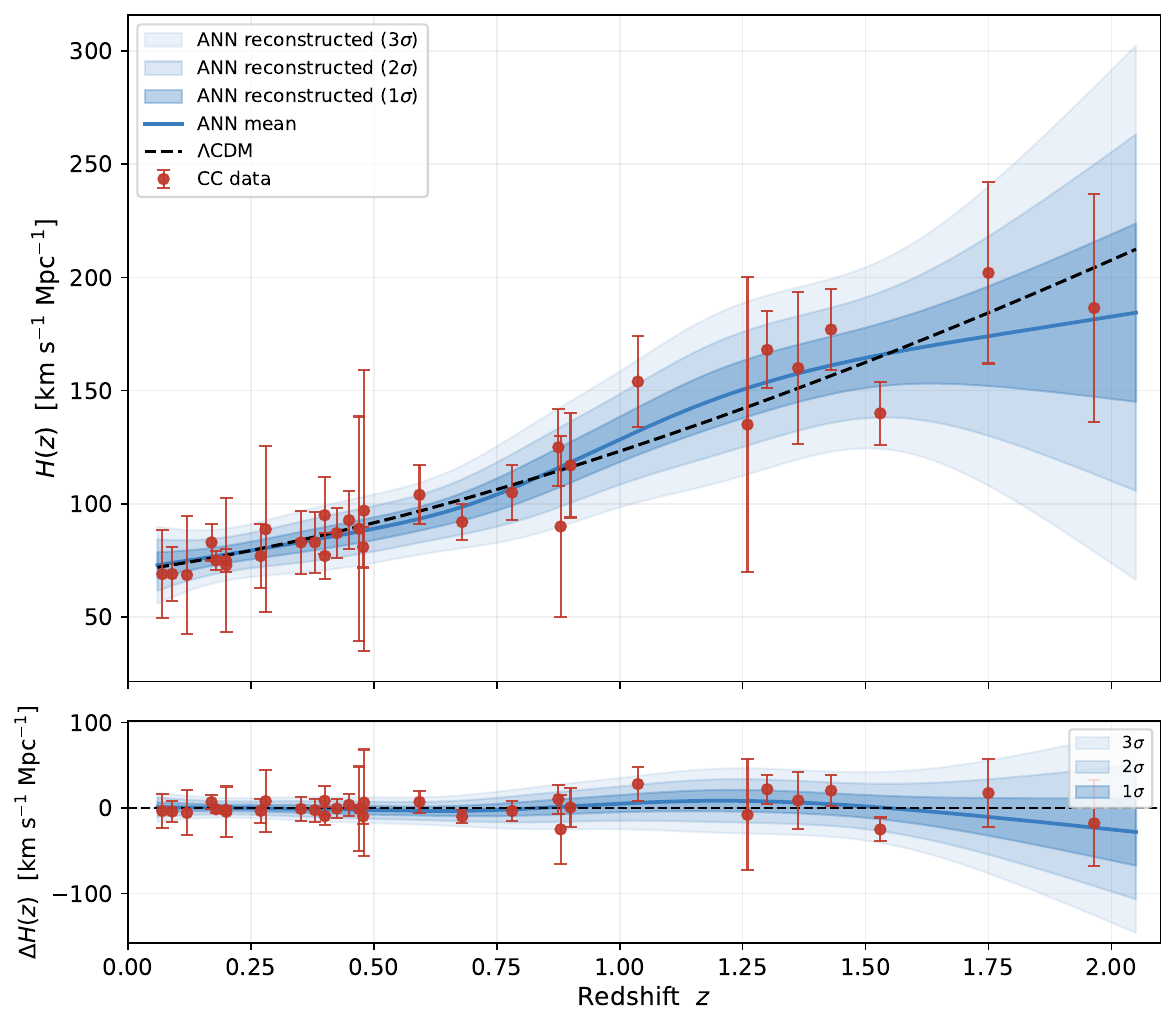}
    \includegraphics[width=0.47\linewidth]{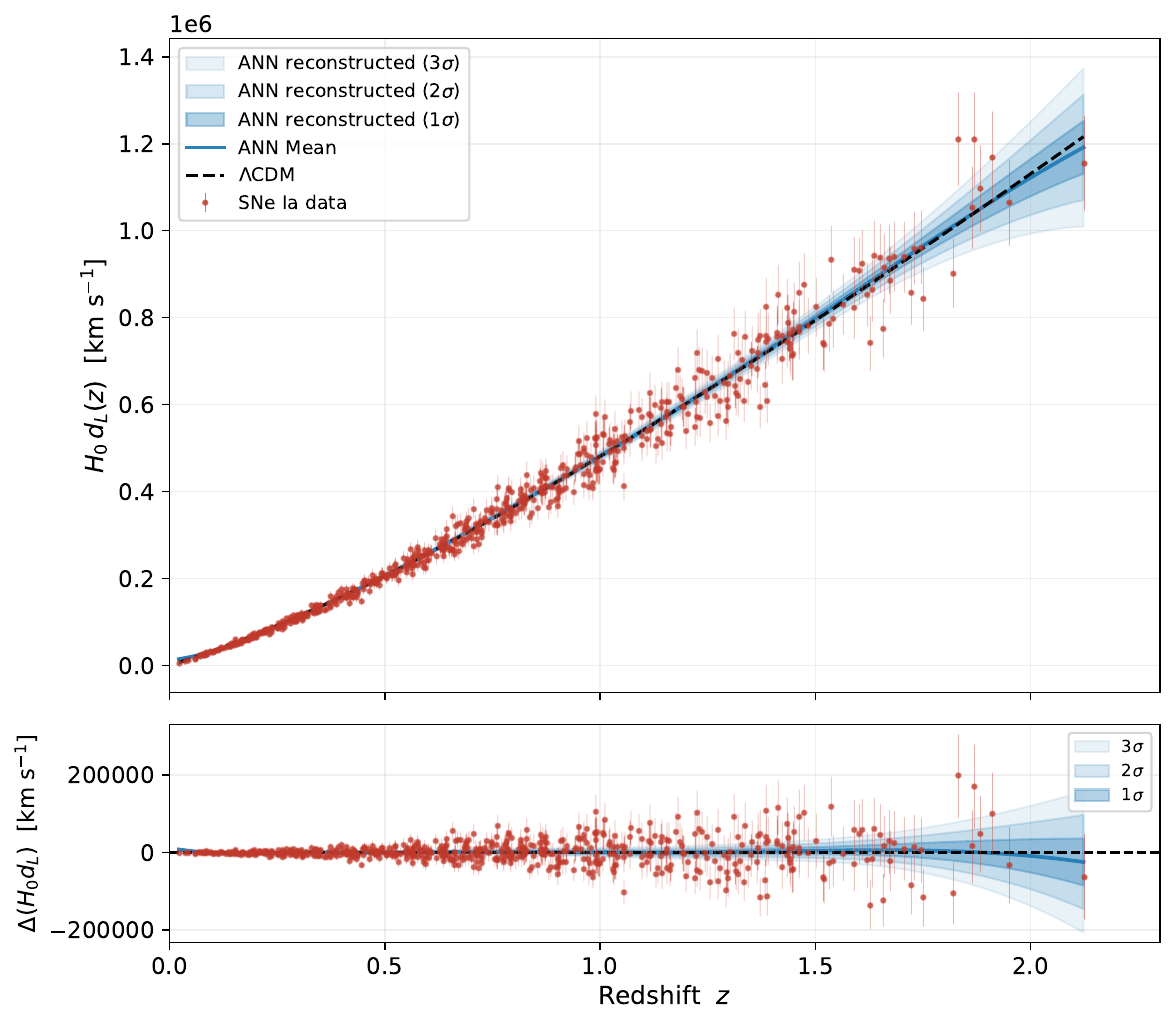}
    \caption{\justifying The left panel shows the reconstruction of the Hubble parameter $H(z)$ from 32 CC measurements and the right panel shows the reconstruction of the unanchored luminosity distance $\Xi_{\rm SNIa}(z) = H_0\,d_L(z)$ from the Pantheon$+$ Type Ia supernova sample using the ANN with 1000 bootstrap realisations. The blue shaded band shows the $1\sigma$ uncertainty of the ANN reconstruction and the blue solid line shows the ANN mean. The red points with error bars are the observational data. }
    \label{fig1}
\end{figure*}

Formally, for each realization $k = 1,\ldots,N_{\rm boot}$, a perturbed dataset is constructed by drawing from the measurement uncertainties:
          \begin{equation}
              y_k = y_{\rm obs} + \delta_k, 
              \quad \delta_k \sim \mathcal{N}(0,\,\sigma_{\rm obs}),
              \label{eq:bootstrap_perturb}
          \end{equation}
 and a fresh ANN model is independently trained on each 
perturbed dataset $y_k$ and evaluated on a fine redshift grid of 600 uniformly spaced points. After all $N_{\rm boot}$ realisations, the reconstruction mean and $1\sigma$ uncertainty at each redshift are computed as,
\begin{equation}
    \bar{f}(z) = \frac{1}{N_{\rm boot}}\sum_{k=1}^{N_{\rm boot}} f_k(z),
    \label{eq:fmean}
\end{equation}
\begin{equation}
    \sigma_{\bar{f}} = \sqrt{\frac{1}{N_{\rm boot}-1}
    \sum_{k=1}^{N_{\rm boot}}\bigl[f_k(z)-\bar{f}(z)\bigr]^2},
    \label{eq:fstd}
\end{equation}
where $f_k(z)$ denotes $H_k(z)$ for the CC reconstruction and 
$\log_{10}(\Xi_{\rm SNIa}^{(k)})$ for the SNe\,Ia reconstruction. This procedure is applied separately to the CC $H(z)$ and SNe\,Ia $\log_{10}(\Xi_{\rm SNIa})$ reconstructions. The spread of the bootstrap ensemble naturally captures how measurement uncertainties propagate through the nonlinear ANN mapping, without assuming any particular shape for the reconstructed function or its uncertainty. The results of the reconstructions for 1000 bootstrap realisations are presented in Fig.~\ref{fig1}.

\section{$H_0$ Posterior Construction}
\label{sec:H0}
\paragraph{Individual Posteriors at Each BAO Redshift:}
At each BAO effective redshift $z_j$ ($j = 1,\ldots,7$), we construct an $H_0$ posterior as follows. The $N_{\rm boot}$ bootstrap curves obtained from the ANN reconstruction are first interpolated to $z_j$, yielding $N_{\rm boot}$ values each of $\Xi_{\rm SNIa}^{(k)}(z_j)$ and $H_{\rm CC}^{(k)}(z_j)$. Simultaneously, an independent set of $N_{\rm boot}$ samples of $\Xi_{\rm BAO}(z_j)$ is drawn from its Gaussian measurement distribution,
\begin{equation}
    \Xi_{\rm BAO}^{(k)}(z_j) 
    \sim \mathcal{N}\!\left(\Xi_{\rm BAO}(z_j),\,
    \sigma_{\Xi_{\rm BAO}}(z_j)\right),
    \label{eq:xi_bao_mc}
\end{equation}
which accounts for the observational uncertainty reported by 
DESI DR2. For each sample $k$, the Hubble constant is then evaluated directly from Eq.~(\ref{eq:h0}),
\begin{equation}
    H_0^{(k)} = \frac{\Xi_{\rm SNIa}^{(k)}\cdot 
    H_{\rm CC}^{(k)}}{(1+z_j)^2\cdot\Xi_{\rm BAO}^{(k)}},
    \label{eq:h0_sample}
\end{equation}
combining the three independently reconstructed quantities at 
the same redshift. A physical sanity cut of $40 < H_0 < 120$\,km\,s$^{-1}$\,Mpc$^{-1}$ is applied to remove any 
unphysical outliers that may arise from rare extreme bootstrap 
realisations. The individual $H_0$ posterior at $z_j$ is then estimated from the surviving samples using a Gaussian kernel density estimator (KDE) with Silverman bandwidth selection, which makes no assumption about the shape of the distribution and naturally captures any asymmetry or non-Gaussianity in the posterior. The central value and asymmetric $1\sigma$ uncertainties are finally reported as the median and the 16th and 84th percentiles of the sample distribution. 
\paragraph{The Joint Constraint:} The seven individual $H_0$ posteriors $P_j(H_0)$ are statistically independent because they are evaluated at different redshifts using independent BAO measurements. Their product therefore gives the joint 
posterior:
\begin{equation}
    P_{\rm joint}(H_0) \propto \prod_{j=1}^{7} P_j(H_0).
    \label{eq:joint}
\end{equation}
The joint posterior is evaluated on a grid of 3000 points 
spanning $H_0 \in [50,\,90]$\,km\,s$^{-1}$\,Mpc$^{-1}$, normalised by numerical integration, and its median and 68\% credible interval are extracted from the cumulative distribution function.

\begin{figure}
    \centering
    \includegraphics[width=1\linewidth]{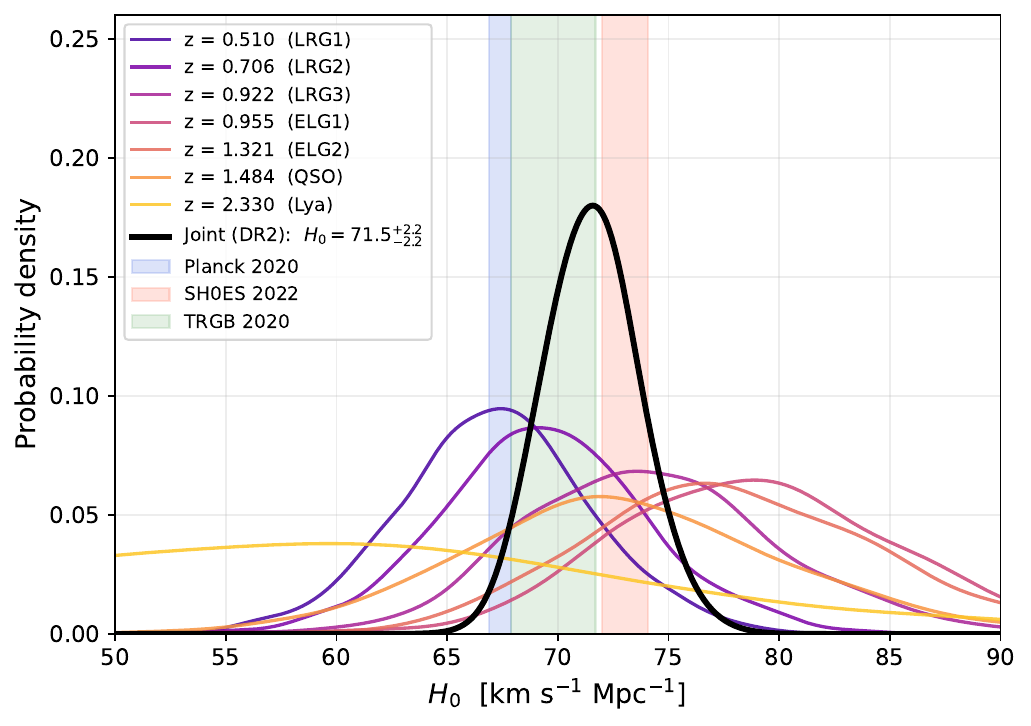}
    \caption{\justifying Posterior probability distribution functions (PDFs) of the Hubble constant $H_0$ at each of the seven DESI DR2 BAO redshifts (coloured curves), together with the joint posterior obtained by multiplying all seven PDFs (black solid curve).} \label{H0}
\end{figure}

We repeat this procedure independently for two configurations: 
$(N_{\rm boot},\,N_{\rm iter},\,N_{\rm neurons}) = (500,\,30000,\,4096)$ 
and $(N_{\rm boot},\,N_{\rm iter},\,N_{\rm neurons}) = (1000,\,30000,\,4096)$, where $N_{\rm boot}$ is the number of bootstrap realisations, $N_{\rm iter}$ is the number of training iterations, and $N_{\rm neurons}$ is the number of neurons in the hidden layer\footnote{Due to the computational cost of training 1000 independent ANN 
models each for 30\,000 iterations with 4096 neurons, these 
calculations were performed on a high-performance workstation 
equipped with an AMD EPYC 9124 16-core processor running at 
up to 3.7\,GHz. Processor specifications: 16 cores, 
CPU max frequency 3.7\,GHz, L2 cache 16\,MiB, L3 cache 
64\,MiB, 64-bit x86 architecture.}. This allows us to verify the stability of our results with respect to the number of bootstrap realisations while keeping the network architecture and training procedure fixed. The two configurations yield consistent results within their uncertainties, as detailed in Table~\ref{tab:H0_results} and the individual $H_0$ posterior PDFs at each of the seven BAO redshifts, together with the joint posterior, obtained from the $N_{\rm boot} = 1000$ configuration are displayed in Fig.~\ref{H0}.






\section{Results and Discussions}\label{section 4}
\begin{figure}
    \centering
    \includegraphics[width=1.0\linewidth]{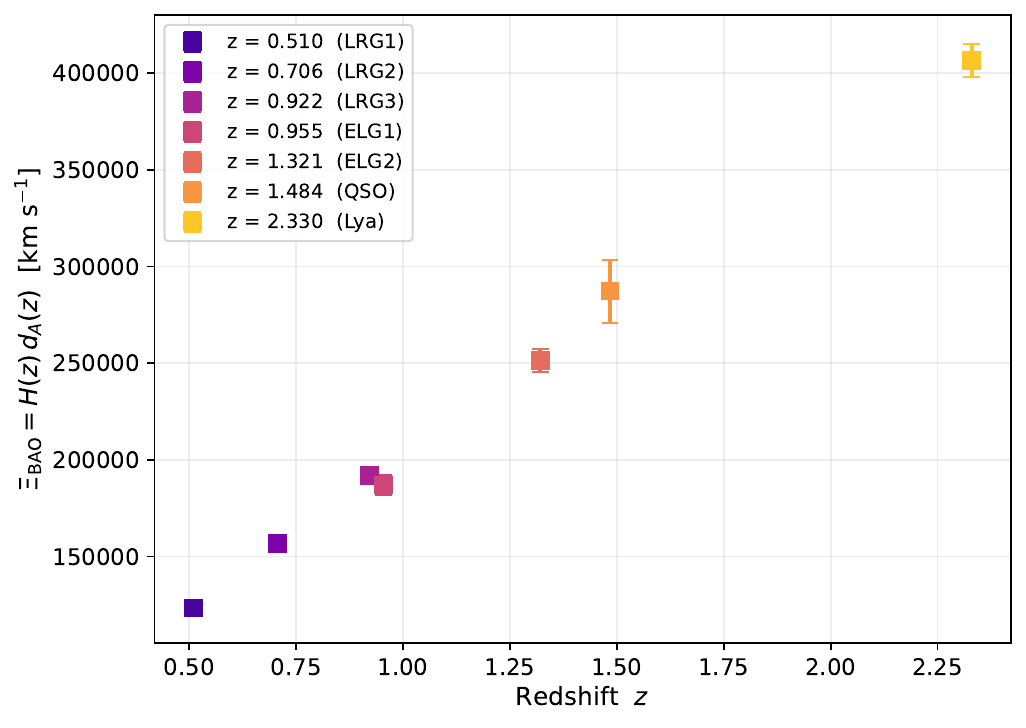}
    \caption{\justifying Sound-horizon-free BAO combination 
$\Xi_{\rm BAO}(z) = H(z)\,d_A(z)$ computed from the seven DESI DR2 BAO measurements at their respective effective redshifts.}
    \label{fig2}
\end{figure}

Our core idea is to measure the sound-horizon-free Hubble constant  $H_0$ using the latest BAO measurements from DESI DR2 \cite{Karim/2025a,Karim/2025b} via the ANN reconstruction method. Our approach is based on the distance duality relation (DDR), which allows us to direct measurement of the Hubble constant without requiring any knowledge of the sound horizon scale $r_d$ because the 
quantity $\Xi_{\rm BAO}$ is constructed from the ratio $D_M/D_H$ so that the sound horizon scale $r_d$ cancels exactly and eliminates any dependence on CMB-based calibration.
Furthermore, we combine it with two more complementary observational probes through DDR, such as the unanchored luminosity distance $\Xi_{\rm SNIa}(z) = H_0\,d_L(z)$ relation and $H_{\rm CC}(z)$ relation using available compilations of SNe Ia data and CC, respectively. Before computing $H_0$, we reconstruct the Hubble function $H_{\rm CC}(z)$ and the unanchored luminosity distance $\Xi_{\rm SNIa}(z) = H_0\,d_L(z)$ using 32 CC data and Pantheon+ samples, respectively, using the ANN with 1000 bootstrap realisations. We can see it in Fig.~\ref{fig1}, which is also consistent with the fiducial $\Lambda$CDM model ($H_0=70$ and $\Omega_{\rm m0}=0.3$) within $1\sigma$ at all redshifts. Note that here we just compare our results with the fiducial $\Lambda$CDM model without adopting it as a prior in our analysis. Both reconstructions confirm that the ANN method provides reliable and smooth function estimates from the available data. Further, we computed the sound-horizon-free BAO combination $\Xi_{\rm BAO}=H(z)d_A(z)$ from the seven DESI DR2 BAO measurements at their respective effective redshifts shown in Fig.~\ref{fig2}. The values increase monotonically with redshift as expected from the expansion history of the universe, and the uncertainties are relatively small for the LRG and ELG tracers at intermediate redshifts, becoming larger for the QSO and Ly$\alpha$ tracers at higher redshifts where the signal-to-noise ratio of the BAO measurement decreases.

By using all reconstruction stuff and the $H_0$ posterior construction method mentioned in Sec.~\ref{sec:H0}, we obtained the individual $H_0$ posterior PDFs at each of the seven DESI DR2 BAO redshifts from the $N_{\rm boot} = 1000$ configuration, which are shown in Fig.~\ref{H0}. The coloured curves represent the individual tracer posteriors, and the black curve shows the joint constraint obtained by multiplying all seven PDFs. We repeat the process independently for the 500 and 1000 bootstrap configurations to verify the stability of our results with respect to the number of bootstrap realisations while keeping the network architecture and training procedure fixed. The two configurations yield consistent results within their uncertainties, as summarised in Table~\ref{tab:H0_results}.

\begin{figure}
    \centering
    \includegraphics[width=1.0\linewidth]{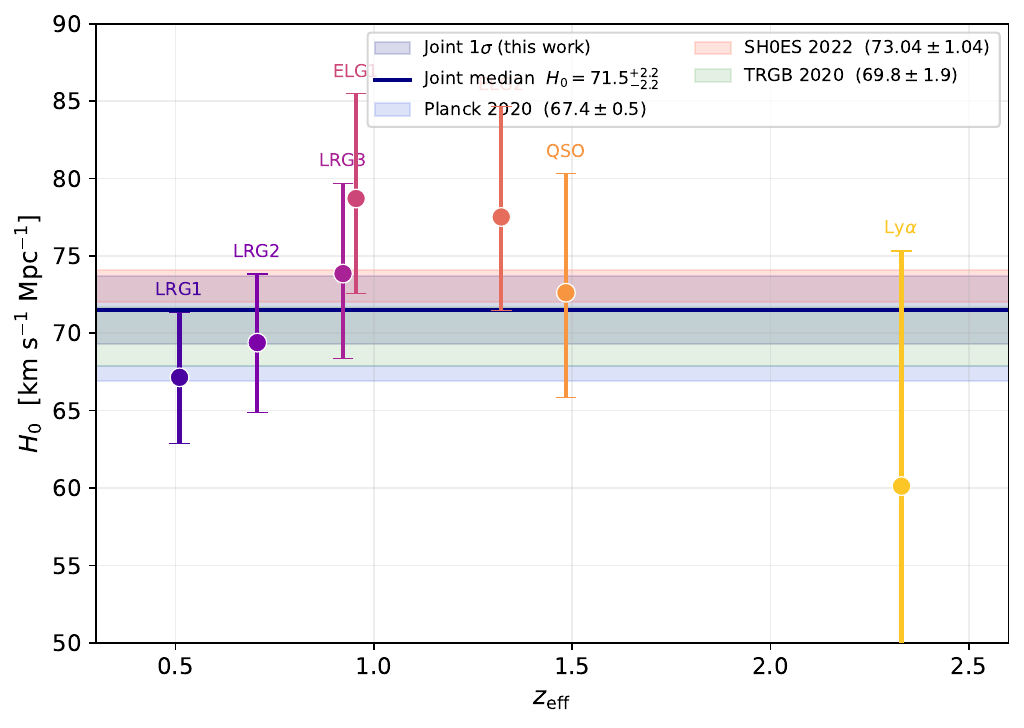}
    \caption{\justifying Hubble constant $H_0$ as a function of effective redshift $z_{\rm eff}$ for each of the seven DESI DR2 BAO tracers (coloured circles). }
    \label{fig3}
\end{figure}

Several features of the individual posteriors are worth noting. For instance, the LRG1 tracer at $z = 0.510$ gives the lowest individual $H_0$ estimate, whereas the ELG1 tracer at $z = 0.955$ and the ELG2 tracer at $z = 1.321$ give the highest. This spread is illustrated in Fig.~\ref{fig3}, which displays $H_0$ as a function of effective redshift. Importantly, the individual estimates are broadly consistent with one another within their relatively large uncertainties. This consistency suggests no strong evidence for a redshift evolution of $H_0$ across the range $0.510 \leq z \leq 2.330$. Such an agreement provides a useful test of the cosmological principle -- if $H_0$ varied significantly with redshift, it would suggest either an inhomogeneous expansion history or systematic effects in one or more of the datasets. It is worth noting that since $H_0$ is an integration constant in the FLRW framework \cite{Krishnan2021}, any statistically 
significant redshift variation of $H_0$ in a model-independent analysis would constitute direct evidence for a breakdown of the cosmological principle, independent of any 
assumed cosmological model \cite{Krishnan2020}. In this context, the marginally increasing-decreasing trend in $H_0$ visible in Figure~\ref{fig3} is consistent with the decreasing-increasing trend in $\Omega_m$ reported in \cite{Colgain2025a}, reflecting the well-known negative correlation between $H_0$ and $\Omega_m$ in flat $\Lambda$CDM, and suggests that tomographic analyses of 
$H_0$ across redshift offer a valuable consistency check on the cosmological principle \cite{Krishnan2020}. The Ly$\alpha$ tracer at $z = 2.330$ lies beyond the Pantheon$+$ SNe\,Ia coverage, so the SNe\,Ia reconstruction at this redshift involves mild extrapolation. As a result, the Ly$\alpha$ posterior is broader, and its central value is somewhat lower than the other tracers.

\begin{figure}
    \centering
    \includegraphics[width=1.0\linewidth]{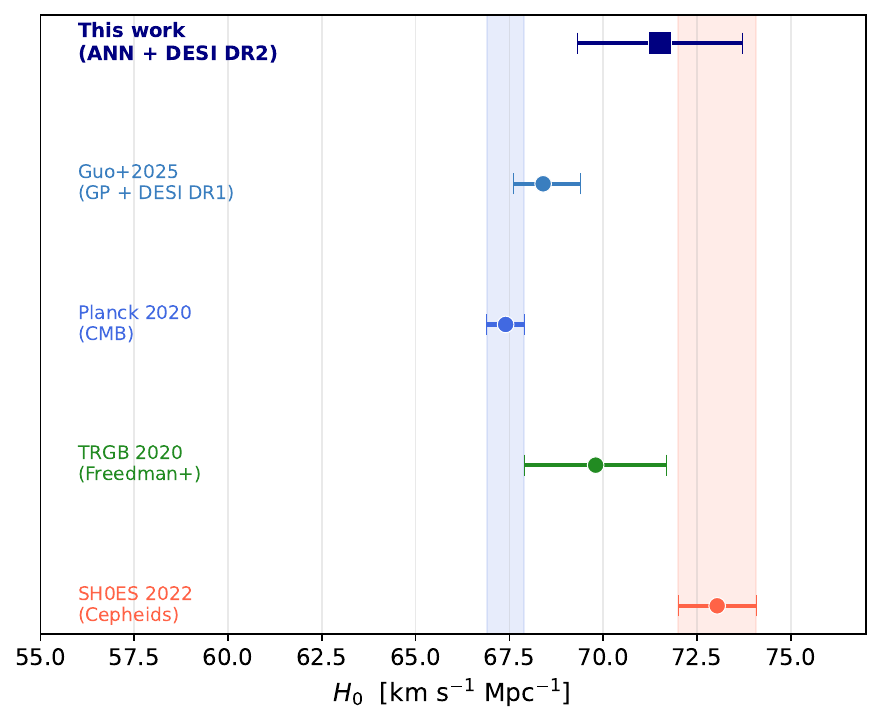}
    \caption{\justifying Comparison of the joint $H_0$ constraint from this work (navy square) with selected results from the literature. Error bars represent $1\sigma$ uncertainties.}
    \label{fig4}
\end{figure}

Multiplying all seven individual posteriors, we obtain the joint constraint: $H_0 = 71.8^{+2.3}_{-2.1}$ $\text{km\,s}^{-1}\text{\,Mpc}^{-1}$ for 500 bootstrap realisations and $H_0 = 71.5^{+2.2}_{-2.2}$ 
$\text{km\,s}^{-1}\text{\,Mpc}^{-1}$ for 1000 bootstrap realisations at the 68\% confidence level.
We compare our joint $H_0$ result with selected results from the literature, which we can see in Fig.~\ref{fig4}. Our result is consistent with the SH0ES and the TRGB measurements within $0.6\sigma$ tension. With respect to the Planck 2020 CMB result, our measurement shows a $\sim$$2\sigma$ tension. Compared with the GP-based DR1 result of 
\cite{Guo2025}, our value is higher by 
$\sim$$1.3\sigma$, which we attribute to three 
sources. First, we use the ANN bootstrap 
method \cite{Wang2020}, which is completely 
data-driven and requires no prior mean function, 
whereas \cite{Guo2025} adopt GP regression with 
the $\Lambda$CDM model as the prior mean function, 
which introduces a mild model dependence that 
could bias the reconstruction toward lower $H_0$ 
values consistent with Planck. Second, we use 
seven anisotropic DESI DR2 tracers compared to 
the five DR1 tracers used in \cite{Guo2025}, 
with the two new tracers ELG1 and QSO 
individually preferring higher $H_0$ values as 
seen in Table~\ref{tab:H0_results}, which 
naturally shifts the joint constraint upward. 
These two effects are likely the main reasons 
for the difference. Third, we propagate the full $\sigma_{a_B}$ calibration uncertainty through the SNe\,Ia reconstruction, while \cite{Guo2025} incorporated $\sigma_{a_B}$ through the full covariance matrix of the GP regression, in our approach we propagate it directly and analytically into the SNe\,Ia uncertainty at the data level via Eq.~(\ref{eq:sigma_xi_snia}), before the reconstruction begins, ensuring that it is fully sampled across all $N_{\rm boot}$ bootstrap realisations and naturally 
propagated into the final $H_0$ posterior.

\begin{table*}[t]
\begin{center}
\begin{minipage}{\linewidth}
\caption{\justifying The Hubble constant $H_0$ (km\,s$^{-1}$\,Mpc$^{-1}$) recovered at each DESI DR2 BAO redshift bin using the ANN reconstruction \cite{Wang2020} with 500 and 1000 bootstrap realisations, compared with the GP-based DR1 results in~\cite{Guo2025}. The row labeled Joint$(H_0)$ gives the combined constraint obtained by multiplying all seven individual posteriors at the 68\% confidence level.}
\label{tab:H0_results}
\end{minipage}
\vspace{0.3em}
\begin{tabular}{l c c c}
\hline\hline
\\[-1.8ex]
Tracer & $H_0$ (500 bootstrap) & $H_0$ (1000 bootstrap) & GP results, DESI DR1 BAO  \\[1ex]
\hline
\\[-1.8ex]
$z = 0.510$ (LRG1)            & $67.3^{+3.9}_{-4.3}$ & $67.2^{+4.2}_{-4.3}$ & $63.4^{+2.3}_{-2.1}$ \\[1ex]
$z = 0.706$ (LRG2)            & $69.3^{+4.9}_{-4.5}$ & $69.4^{+4.4}_{-4.5}$ & $72.4^{+2.8}_{-2.5}$ \\[1ex]
$z = 0.922$ (LRG3)            & $73.8^{+5.6}_{-5.0}$ & $73.9^{+5.8}_{-5.5}$ & $70.8^{+1.8}_{-1.8}$ \\[1ex]
$z = 0.955$ (ELG1)$^{\color{red}{a}}$      & $78.2^{+7.3}_{-5.0}$ & $78.7^{+6.8}_{-6.1}$ & \multicolumn{1}{c}{---} \\[1ex]
$z = 1.321$ (ELG2)            & $77.7^{+6.5}_{-6.0}$ & $77.5^{+7.1}_{-6.0}$ & $67.9^{+2.9}_{-2.4}$ \\[1ex]
$z = 1.484$ (QSO)$^{\color{red}{a}}$       & $72.3^{+7.4}_{-6.5}$ & $72.6^{+7.7}_{-6.8}$ & \multicolumn{1}{c}{---} \\[1ex]
$z = 2.330$ (Ly$\alpha$)$^{\color{red}{b}}$ & $59.2^{+14.9}_{-11.1}$ & $60.1^{+15.2}_{-12.7}$ & $67.5^{+2.3}_{-2.3}$ \\[1ex]
\hline
\\[-1.8ex]
Joint$(H_0)$                  & $71.8^{+2.3}_{-2.1}$ & $71.5^{+2.2}_{-2.2}$ & $68.4^{+1.0}_{-0.8}$ \\[1ex]
\hline\hline
\\[-1.8ex]
\multicolumn{4}{l}{$^{\color{red}{a}}$ DESI DR1 did not include ELG1 ($z=0.955$) and QSO ($z=1.484$) as separate anisotropic tracers;} \\
\multicolumn{4}{l}{\phantom{$^{\color{red}{a}}$}  these redshift bins are new in DR2.} \\[0.5ex]
\multicolumn{4}{l}{$^{\color{red}{b}}$ The Ly$\alpha$ redshift ($z=2.330$) exceeds the Pantheon$+$ SNe\,Ia coverage ($z \lesssim 2.26$); the reconstruction} \\
\multicolumn{4}{l}{\phantom{$^{\color{red}{b}}$}  relies on mild extrapolation, which may introduce a small bias.} \\
\end{tabular}
\end{center}
\end{table*}
\section{Conclusions}\label{section 5}
In this work, we obtained a model-independent result, free from any sound-horizon calibration and any SNe Ia luminosity assumption, that favors a higher value of $H_0$ closer to the SH0ES and TRGB measurements than the Planck CMB inference. The key advantage of our approach is that it relies solely on directly observable quantities, such as the unanchored luminosity distance from SNe Ia, the Hubble parameter from CC, and the sound-horizon-free BAO combination from DESI DR2 -- without invoking any cosmological model at any stage of the analysis. This makes our result a truly independent data point in the Hubble tension debate, free from the model-dependent assumptions that affect both the Planck CMB inference and the local distance ladder. The consistency 
of our results across both the 500 and 1000 bootstrap realisations confirms the internal stability and robustness of the ANN reconstruction method.

Moreover, several factors contribute to the relatively large uncertainties in our individual tracer estimates compared with the GP-based results in~\cite{Guo2025}. First, the bootstrap ANN procedure introduces additional variance from the random weight initialization and the stochastic training process in each realization. Second, the two new DR2 tracers, such as ELG1 and QSO, have relatively larger BAO measurement uncertainties compared to the LRG tracers, which broadens their individual posteriors. Despite these larger individual uncertainties, the joint constraint benefits from the addition of two new independent redshift bins compared to DR1, demonstrating the value of the expanded DESI DR2 dataset. Third, we propagate the calibration uncertainty $\sigma_{a_B}$ 
analytically at the data level before the reconstruction begins, whereas \cite{Guo2025} incorporate it through the GP covariance matrix, which enters the uncertainty at a different stage of the analysis.

Looking ahead, the forthcoming DESI DR3 and the final five-year dataset, which will cover the full 14,200 square degrees of the survey footprint across a wider redshift range, will provide BAO measurements of significantly higher precision than the current DR2, and combined with our framework are expected to reduce the joint $H_0$ uncertainty to the sub-percent level. Furthermore, the Vera Rubin Observatory LSST \cite{LSST2019} will extend the SNe\,Ia coverage to $z \sim 3$, eliminating the mild extrapolation currently required at the Ly$\alpha$ redshift ($z = 2.330$) and making all tracers fully reliable. Together, these next-generation datasets will allow our completely model-independent framework to provide a definitive and independent contribution to resolving the Hubble tension in the coming decade.

\acknowledgments
Gaurav N.Gadbail and Kazuharu Bamba appreciate the support by the JSPS KAKENHI Grant No.~25KF0176. The work of Kazuharu Bamba was also supported in part by the JSPS KAKENHI Grant No.~24KF0100 and a grant-in-aid of academic research of the Yamaguchi Scholarship Foundation. In addition, this work was supported from the Research Promotion Project of Fukushima University Fund.


\twocolumngrid

\end{document}